\documentclass[12pt]{iopart}
\usepackage{iopams}
\usepackage{graphicx}
\usepackage{dcolumn}
\usepackage{bm} 
\newcommand{\be}{\begin{equation}}
\newcommand{\ee}{\end{equation}}
\newcommand{\bef}{\begin{figure}}
\newcommand{\eef}{\end{figure}}
\newcommand{\bea}{\begin{eqnarray}}
\newcommand{\eea}{\end{eqnarray}}
\newcommand{\ba}{\begin{array}}
\newcommand{\ea}{\end{array}}
\newcommand{\bds}{{\bf dS}}

\newcommand{\eps}{\epsilon}
\newcommand{\na}{\vec{\nabla}}
\begin{document}

\title{Clustering and coalescence from multiplicative noise: 
the Kraichnan ensemble.} 

\author{Andrea Gabrielli, Fabio Cecconi}
\address{SMC - INFM/CNR, Department of Physics,
University ``La Sapienza'', P.le Aldo Moro 2, 00185-Rome, Italy\\
and ISC - CNR, via dei Taurini 19, 
00185-Rome, Italy}

\begin{abstract}
We study the dynamics of the two-point statistics of the Kraichnan
ensemble which describes the transport of a passive pollutant by a
stochastic turbulent flow characterized by scale invariant structure
functions.  The fundamental equation of this problem consists in the
Fokker-Planck equation for the two-point correlation function of the
density of particles performing spatially correlated Brownian motions
with scale invariant correlations. This problem is equivalent to the
stochastic motion of an effective particle driven by a generic
multiplicative noise.  In this paper we propose an alternative and
more intuitive approach to the problem than the original one
\cite{verg-gaw} leading to the same conclusions.  The general features
of this new approach make possible to fit it to other more complex
contexts.
\end{abstract}


\maketitle 

\section{Introduction} \label{intro}
Correlated random walks have always received interest from different
scientific disciplines thanks to their ability to generate complex
patterns and collective behaviors
\cite{appl1,appl2,appl3,appl4,appl5,lopez,cecco}.  A case of
particular relevance amounts to considering the case of a particle
distribution in which single particles perform an ordinary Brownian
motion, but the trajectories of different particles are spatially
correlated \cite{displa,nostro1}.  A rich phenomenology occurs in this
case ranging from simple diffusion, to clustering and coalescence of
particles \cite{mehlig,Deutsch}.  The so called Kraichnan ensemble
\cite{kraich,falkovich-rev}, introduced to describe the transport of
passive scalars by turbulent flows, belongs to this class of
models. In this approach, the turbulent flow is modeled as a
stochastic velocity field with no time correlations (i.e., white in
time) and scale invariant spatial correlations.  Passive particles
(e.g. pollutant) are considered to have vanishing mass (no inertia),
and are simply transported by the flow without affecting it.  As we
discuss below, from the point of view of non-equilibrium statistical
physics, this ensemble of models is equivalent to the general problem
of a $d-$dimensional Langevin equation driven by multiplicative noise
only. The whole ensemble has been solved in \cite{verg-gaw,gaw} using
the mathematical theory of boundary conditions of elliptic
differential operators developed in rigorous quantum mechanics
\cite{reed-simon}.

In this paper, besides showing the strict relationship between
Kraichnan turbulence and the physics of multiplicative noise, we
present an alternative approach to these models leading to the same
quantitative classification of the phases occurring in the Kraichnan
ensemble. Our approach deals directly with the equation for the two-point 
correlation function of the density of pollutant instead of the correlation
function of the passive scalar.  
The two equations are known to be equivalent being one
the adjoint of the other. The reason to propose this alternative
solution is twofold: (i) it represents a more intuitive technique
based on a simple regularization of the equation at small scales. This
regularization amounts to studying the integrated mass of pollutant
surrounding a generic pollutant particle.  (ii) The new method is so
general that can be potentially applied to more complex cases which,
up to now, have been solved only under particular conditions
\cite{Cencio,Cencio1}. 

The key point of the present method is an ``exponent hunter'' technique 
allowing an exact classification of particle-particle correlations 
at vanishing interparticle distance where the dynamics presents a singularity. 

The equation we study describes three  
different regimes for the pollutant concentration   
which depend on: space dimensionality, anisotropy and scaling properties 
of the velocity field correlations. 
These regimes correspond to three different
boundary like situations for the singularity of the equation at small
interparticle separation: 
\begin{enumerate}
\item {\em Phase of strong flow compressibility}.  There is only one
possible solution of the problem, as different particles collide with
vanishing relative velocity. Thus particles coalesce in finite time
and no stationary state is approached by the system;
\item {\em Phase of weak compressibility}. Particles never collide and
the system converges towards a stationary state characterized by power
law or ``fractal-like'' density correlations;
\item {\em Phase of intermediate compressibility} (called also {\em
sticky} phase \cite{gaw}). Particles collide in finite time with
non-zero relative velocity.  Therefore both previous cases are
possible depending on the boundary condition, absorbing or reflecting,
assigned ``by hand'' at vanishing interparticle separation.
\end{enumerate} 
Note that for the first two cases there is no chance to 
impose arbitrary boundary condition at vanishing interparticle
separation because the solution of the fundamental equation is
unique.

The paper is structured as follows: in sec.~\ref{basic} we briefly review  
the basic properties of the Kraichnan ensemble. 
In sec.~\ref{hunter} we discuss the method of solution of the 
Fokker-Planck like equation describing the evolution of the two point
correlation function of the pollutant concentration. 
Finally sec.~\ref{fine} is devoted to discussion and conclusions.

\section{The Kraichnan ensemble and basic equations} \label{basic}
We start the discussion by introducing briefly, in the context of
stochastic processes, the definition of the Kraichnan ensemble.  The
microscopic concentration (or particle density) of the $N$ passive
pollutant particles in the $d-$dimensional volume $V$ is
$$
\rho(\bm x,t)=\sum_{i=1}^N\delta[\bm x-\bm x_i(t)]
$$
in which $\bm x_i(t)$ represents the position at time $t$ of the $i^{th}$ 
pollutant particle transported by the turbulent flow.
The particle distribution is supposed to be spatially homogeneous 
so that $\left<\rho(\bm x,t)\right>=\rho_0=\lim_{V\to \infty} N/V >0$.
Each particle is considered to have a negligible mass,
and consequently it will follow the trajectory defined by the Lagrangian flow:
\be
\frac{d\bm x(t)}{dt}=\bm v[\bm x(t),t]\,,
\label{eq-r}
\ee
where $\bm v(\bm x,t)$ is the Gaussian stochastic velocity field representing
the ``synthetic'' turbulent flow advecting the pollutant.
The spatio-temporal correlation properties of this field are taken to be
\be
\left\{
\begin{array}{l}
\overline{\bm v(\bm x,t)}=0\\
\overline{v_\mu(\bm x,t)v_\nu(\bm x',t')}=\delta(t-t')c_{\mu\nu}(\bm x-\bm x')
\end{array}
\right.\;,
\label{corr-v}
\ee 
with $\mu,\nu=1,...,d$. In other words, the velocity field is considered white
in time and colored in space.  

We limit our study to the description of the two-particle dynamics and 
correlations. The higher order statistics has been characterized in 
Ref.~\cite{verg-gaw}.    
The relative vector distance $\bm r(t)=\bm x_i(r)-\bm x_j(t)$ between any 
pair of particles satisfies the closed Langevin equation 
in \^Ito representation\footnote{A simple way
to understand why Eq.~\ref{eq-r} has to be interpreted in Ito rather than 
Stratonovich representation 
is that the corresponding Fokker-Planck equation must not contain the drift 
term, as the stochastic advecting flow is considered isotropic.}:
\be
\frac{d\bm r(t)}{dt} = \bm w(t)\,.
\label{eq-Dr}
\ee 
The quantity $\bm w(t) = \bm v[\bm x_i(t),t] - \bm v[\bm x_j(t),t]$ is a
Gaussian noise whose correlation properties are 
\be 
\left\{
\begin{array}{l}
\overline{\bm w(t)}=0\\
\overline{w_\mu(t)w_\nu(t')}=2\delta(t-t')d_{\mu\nu}[\bm r(t)]
\end{array}
\right.
\label{corr-w}
\ee 
where $d_{\mu\nu}(\bm r)=[c_{\mu\nu}(0)-c_{\mu\nu}(\bm r)]$ is the so-called
{\em structure tensor} of the velocity field.
Equations (\ref{eq-Dr}) and (\ref{corr-w}) indicate that 
the quantity $\bm r(t)$ satisfies the most general Langevin 
equation for multiplicative noise \cite{multiplicative,Munoz}.

The Kraichnan ensemble is defined by the scale invariant choice
\be 
d_{\mu\nu}(\bm r)=a r^\xi\delta_{\mu\nu}+br^{\xi-2}r_\mu r_\nu
\label{eq2}
\ee
with $a$ and $b$ constant such that $c_{\mu\nu}(\bm r)$ is a non-negative 
definite rank $2$ tensor. This means that 
its Fourier transform (power spectrum tensor)
$$
\tilde c_{\mu\nu}(\bm k)=\int d^dr\, c_{\mu\nu}(\bm r) e^{-i\bm k\cdot\bm r}
$$
has non-negative eigenvalues for all $\bm k$.  
Such a condition implies also that $0<\xi\le 2$ \cite{nostro1}.  
The one-dimensional case is recovered when setting 
$d=1$ and $b=0$. 
In the paper \cite{verg-gaw} and all related literature, 
the tensor $d_{\mu\nu}(\bm r)$ is written in
terms of two other constants, $A$ and $B$, such that 
\be 
\left\{
\begin{array}{l}
a = A + (d+\xi-1)B\\
b = \xi(A-B)
\end{array}
\right.\;.
\label{constants}
\ee 
The positive definiteness of the covariance tensor
$c_{\mu\nu}(\bm r)$ is guaranteed by taking $A,B\ge 0$, which is
equivalent to the condition $-(\xi a)/(d+\xi-1)\le b\le\xi a$.  These
two constants are introduced as $A = 0$ corresponds to the
incompressible case where $\nabla\cdot\bm v = 0$ and $B = 0$ to the
purely potential one with $\bm v = \nabla \phi$.  The quantities ${\cal
S}^2 = A+(d-1)B$ and ${\cal C}^2 = A$ are proportional to
$\left<\|\nabla\bm v\|^2\right>$ and $\left<(\nabla\cdot\bm v)^2\right>$
respectively, and they satisfy the inequalities ${\cal S}^2\ge {\cal
C}^2 \ge 0$.  The degree of compressibility is defined as the ratio
${\cal P}={\cal C}^2/{\cal S}^2\in [0,1]$.  The bounds ${\cal P}=0$
and ${\cal P}=1$ define a completely incompressible and
a completely compressible flow respectively.  In one dimension, ${\cal
S}^2= {\cal C}^2\ge 0$ and therefore ${\cal P}=1$. Clearly, in real
flows Eq.~(\ref{eq2}) can hold only in a limited range of scales, as
in general $c_{\mu\nu}(\bm r)$ must vanish for $r\to \infty$, and
accordingly $d_{\mu\nu}(\bm r)$ must converge to the constant
$c_{\mu\nu}(0)$.

As a consequence of Eq.~(\ref{eq2}), the point $\bm r=0$ is a stationary point 
of Eq.~(\ref{eq-Dr}). As shown below, it may determine, for particular values 
of the parameters ${\cal P}$, $d$ and $\xi$, an absorbing phase for the 
dynamics. We will discuss this problem in the Fokker-Planck formalism. 

The Langevin equation (\ref{eq-Dr}) is equivalent \cite{Gardiner} 
to the $d-$dimensional Fokker-Planck equation (FPE) 
\be 
\partial_t P(\bm r,t|\bm r_0,t_0)=\sum_{\mu,\nu}^{1,d}\partial^2_{\mu\nu}
\left[d_{\mu\nu}(\bm r) P(\bm r,t|\bm r_0,t_0)\right]\,,
\label{eq1}
\ee 
describing the evolution of the probability density function (PDF)
$P(\bm r,t|\bm r_0,t_0)$ of the interparticle distance $\bm r$ at time $t$ given the
initial value $\bm r_0$ at time $t_0$.  In Eq.~(\ref{eq1})
$\partial_t=\frac{\partial}{\partial t}$ and $\partial^2_{\mu\nu}=
\frac{\partial^2}{\partial r_\mu\partial r_\nu}$ with $r_\mu$ being
the $\mu^{th}$ component of vector $\bm r$.  In turbulence literature
(e.g., see \cite{falkovich-rev,verg-gaw}) instead of Eq.~(\ref{eq1}),
one usually studies the adjoint equation which is called, in the
stochastic processes context, a ``backward'' FPE, while
Eq.~(\ref{eq1}) is called the ``forward'' FPE. They are practically
equivalent, even though the differential operators in the two
equations are defined on different functional spaces \cite{feller55}.
The two equations coincide for incompressible flows.

Equation (\ref{eq1}) is a $d$-dimensional FPE without drift, typical
of generalized multiplicative noise.  If the tensor $d_{\mu\nu}(\bm r)$
never vanishes in the interior of the domain of definition of
Eq.~(\ref{eq1}), a unique solution exists when the initial
[e.g. $P(\bm r,t_0|\bm r_0,t_0)=\delta(\bm r-\bm r_0)$] and the boundary
conditions are given \cite{Gardiner,Vankampen}.  However, if
$d_{\mu\nu}(\bm r)$ vanishes at some point, the solution could be no more
unique and an additional boundary condition might be necessary at such
a singularity. 

We develop a novel method to provide, not only the qualitative
classification of the singularity $\bm r={\bf 0}$ for the Kraichnan ensemble
(see \ref{appa}), but also the exact solution of
Eq.~(\ref{eq1}) in the form of a particular series expansion in the
neighborhood of the singularity. This represents an alternative and
more intuitive approach than the original one \cite{verg-gaw}.   This
method can be potentially generalized and applied to different kind of
FPE with singular points in the domain of definition.

It can happen that the singularity $\bm r={\bf 0}$ behaves as an
additional regular boundary \footnote{We use here the definitions
given by Van Kampen in Chap. XII of its book \cite{Vankampen}. This
classification differs from that given originally by Feller
\cite{feller55} and re-proposed by Gardiner \cite{Gardiner}. The two
are related directly by the following simple positions: ``adhesive
boundary'' becomes ``exit boundary'' in the latter, ``natural
repulsive'' becomes ``entrance'' or ``natural'' depending if the
average time to be expelled at finite distance from the singularity is
respectively finite or infinite, and finally``natural attractive''
also becomes simply ``natural''.}. Then an absorbing, reflecting or
mixed boundary condition needs to be assigned.  In all the other cases
the singularity can be seen as a boundary at which the condition is
unique and automatically fixed by the form of Eq.~(\ref{eq1}) itself.
In particular when the singularity corresponds to an adhesive boundary
(see \ref{appa}), the diffusion of the effective particle
described by Eq.~(\ref{eq1}) is such that the boundary is reached in a
finite time and it behaves automatically as an absorbing boundary
\cite{Vankampen}. Naively one can say that the effective particle,
whose position is $\bm r(t)$, arrives at this boundary in a finite time
but with vanishing velocity obliging it to stay at that point
indefinitely.  Instead when the singularity behaves as a natural
attractive boundary the same effective particle approaches it, but only
in an infinite time. Therefore there is no need of external
boundary conditions at the singularity.  Finally, for a singularity which is a
natural repulsive boundary for Eq.~(\ref{eq1}), the effective
particle, when placed arbitrarily close to this point, is expelled far
from it.  We will see that this last situation, in turn, splits into
two sub-cases in which the time to be expelled at a finite distance is
respectively finite or infinite.  The details about the Van Kampen
classification will be discussed in \ref{appa}.

\section{Method and solution}\label{hunter}
In this paper, instead of considering the PDF $P(\bm r,t|\bm r_0,t_0)$,
we study the equivalent problem of the evolution of the two-point correlation 
function of the pollutant particle density at time $t$
\be
\Gamma(\bm r,t)=\frac{\left<\rho(\bm x,t)\rho(\bm x+\bm r,t)\right>}{\rho_0}
-\delta(\bm r)\,,
\label{gamma}
\ee where we have subtracted the Dirac delta contribution in $\bm r=0$,
to exclude the self-correlation of each particle with itself
\cite{book}. In this way $\Gamma(\bm r,t)$ gives exactly the average
conditional density of other particles seen by a generic particle at
separation $\bm r$ from itself.  The symbol $\left<...\right>$ indicates
the ensemble average over the realizations of the process (the
hypothesis of statistical spatial homogeneity is done implicitly).  It
is noteworthy that this average in Eq.~(\ref{gamma}) implies both the
average $\overline{(...)}$ over the realizations of the
velocity field and the ensemble average over the initial particle
configurations.  From the knowledge of $P(\bm r,t|\bm r_0,t_0)$, we can
write:
$$
\Gamma(\bm r,t)=\int d^dr_0\,\Gamma(\bm r_0,t_0)P(\bm r,t|\bm r_0,t_0)\,.
$$
Consequently,  $\Gamma(\bm r,t)$ satisfies the same forward FPE
as $P(\bm r,t|\bm r_0,t_0)$, which can be cast to the tensorial form:
\be
\partial_t \Gamma(\bm r,t)=\na\cdot\left\{\na\cdot[\hat d(\bm r)
\Gamma(\bm r,t)]\right\}\,,
\label{eq2-1}
\ee
where $\hat d(\bm r)$ is the flow structure tensor which, in intrinsic 
form, reads
\be
\hat d(\bm r)=a r^\xi \hat I+br^{\xi-2}(\bm r\otimes\bm r)\,
\label{eq3}
\ee 
with $\hat I$ being the identity tensor and $\bm r\otimes\bm r$ the tensor
product of $\bm r$ for itself.  Let us now integrate explicitly
Eq.~(\ref{eq2-1}) over the sphere $S(r)$ of radius $r$ around the
origin. The quantity 
\be
N(r,t)=\int_{S(r)}d^dr'\,\Gamma(\bm r',t)=\int_0^r
dr'\,r'^{d-1}\int_{\Omega_T}d\Omega\,\Gamma(\bm r',t)
\label{eq2-2}
\ee
is the number of other particles seen in average by a particle
in the sphere of radius $r$ centered on it, and $\Omega_d$ is the total 
solid angle in $d$ dimensions.
Note that from Eq.~(\ref{eq2-2}) we have 
\be 
\int_{\Omega_d}d\Omega\,\Gamma(\bm r,t)=\frac{1}{r^{d-1}}\partial_r 
N(r,t)\,.
\label{eq2-3}
\ee
Using Eq.~(\ref{eq2-2}) and the divergence theorem of integral calculus,
we can write Eq.~(\ref{eq2-1}) as
\be
\partial_t N(r,t)=\int_{\partial S(r)}\bds\cdot\left\{\na\cdot[\hat d(\bm r)
\Gamma(\bm r,t)]\right\}\,,
\label{eq2-4}
\ee
where $\partial S(r)$ indicates the surface of the sphere $S(r)$, and
$\bds= r^{d-1}d\Omega\,\hat{\i}_r$ is the vector surface element of
the sphere of radius $r$ along the 
radial unitary vector $\hat{\i}_r$. 

In order to transform the integral in Eq.~(\ref{eq2-4}) with the
condition (\ref{eq3}) into a more treatable expression, we apply the
tensor calculus in $d=3$, and then we see how to extend the
description to arbitrary dimensions $d$. The three-dimensional vector
gradient operator in spherical coordinates in the local orthogonal
spherical reference frame, is
$$
\na=\hat{\i}_r \partial_r+\frac{\hat{\i}_{\theta}}{r}\partial_\theta+
\frac{\hat{\i}_{\phi}}{r\sin \theta}\partial_\phi\,.
$$
In the same frame, the tensor product $\bm r\otimes\bm r$ acquires the simple
form
$$
\bm r\otimes\bm r=r^2(\hat{\i}_r\otimes\hat{\i}_r)\,,
$$
while the identity operator is obviously
$$
I=\hat{\i}_r\otimes\hat{\i}_r+\hat{\i}_\theta\otimes\hat{\i}_\theta+
\hat{\i}_\phi\otimes\hat{\i}_\phi\,.
$$
Note that
$\hat{\i}_{r(\theta,\phi)}\otimes\hat{\i}_{r(\theta,\phi)}$ is
the local projector operator along the direction
$\hat{\i}_{r(\theta,\phi)}$.

Using the expression (\ref{eq3}) and the rules of tensor calculus in the
spherical coordinates frame \footnote{In particular we use that any
derivatives of the identity operator $I$ trivially vanishes, that
$\partial_r \hat{\i}_r=0$, $\partial_\theta \hat{\i}_r=\hat{\i}_\theta$
and $\partial_\phi \hat{\i}_r=\sin \theta\,\hat{\i}_\phi$,
and that $\hat{\i}_{r,\theta,\phi}\otimes\hat{\i}_{r,\theta,\phi}$
are projection operators.}, one can write
\bea
&&\na\cdot[\hat d(\bm r)\Gamma(\bm r,t)]=\left[(a+b)\partial_r(r^\xi
\Gamma)+2b\,r^{\xi-1}\Gamma\right]\hat{\i}_r\nonumber\\
&&+a\,r^{\xi-1}\left[
(\partial_\theta\Gamma)\hat{\i}_\theta
+\frac{\partial_\phi\Gamma}{\sin \theta}\hat{\i}_\phi\right]\,,
\label{eq2-5}
\eea
and therefore simply
\be
\hat{\i}_r\cdot\{\na\cdot[\hat d(\bm r)\Gamma(\bm r,t)]\}=
(a+b)\partial_r[r^\xi\Gamma(\bm r,t)]+2b\,r^{\xi-1}
\Gamma(\bm r,t)\,.
\label{eq2-6}
\ee 
Note that the radial direction is the only important spherical
component of Eq.~(\ref{eq2-5}). This is exactly the reason why we have
used the local spherical orthogonal frame.  

In $d$ dimensions
the gradient operator in local hyper-spherical coordinates is
\be
\na=\hat{{\i}}_r\partial_r+\sum_{j=1}^{d-1}\hat{{\i}}_{\psi_j}
\nabla_{\psi_j}\,,
\label{grad}
\ee
where $\nabla_{\psi_j}$ is the component of the gradient along the 
orthogonal angular direction $\hat{{\i}}_{\psi_j}$ on the $d-$dimensional
sphere of radius $r$.
Since the curvature of the sphere is constant and the hyper-spherical 
frame is orthogonal, it is simple to show that for any $j=1,...,d-1$
we have \footnote{Due to the constant curvature of the
sphere and the orthogonality of the coordinate frame, in
order to find $\nabla_{\psi_j}\hat {\i}_r$, it is sufficient to consider the
equatorial circumference in the direction $\hat {\i}_{\psi_j}$ and
therefore the variation of $\hat {\i}_r$ along this. Clearly, as all
equatorial circumferences passing through a point of the spherical
surface are a rotation one of each other, the modulus of 
$\nabla_{\psi_j}\hat {\i}_r$ is the same for any direction $\hat{\i}_{\psi_j}$ 
and depends only on $r$. It is finally simple to find that such a dependence 
is $1/r$.}
$$
\nabla_{\psi_j}\hat{\i}_r=\frac{\hat{\i}_{\psi_j}}{r}
$$
and therefore 
$$
\hat{\i}_{\psi_j}\cdot\nabla_{\psi_j}(\hat{\i}_r\otimes\hat{\i}_r)=
\frac{\hat{\i}_r}{r}\,.
$$ 

This observation permits to generalize the result (\ref{eq2-6}) to any
dimension $d$ in the following way: 
\be 
\hat{\i}_r\cdot\{\na\cdot[\hat
d(\bm r)\Gamma(\bm r,t)]\}=
(a+b)\partial_r[r^\xi\Gamma(\bm r,t)]+(d-1)b\,r^{\xi-1} \Gamma(\bm r,t)\,.
\label{eq2-7}
\ee 
In this relation, the only dependence on the angular variables
lies on $\Gamma(\bm r,t)$. Plugging Eqs.~(\ref{eq2-7}) and (\ref{eq2-3})
into Eq.~(\ref{eq2-4}), we finally arrive at the following closed
equation for $N(r,t)$: 
\bea 
&&\partial_t
N(r,t)=(a+b)r^{d-1}\partial_{r}\left[r^{\xi-d+1} \partial_r
N(r,t)\right]\nonumber\\ &&+(d-1)b\,r^{\xi-1}\partial_r N(r,t)\,.
\label{eq2-8}
\eea 
The singularity $\bm r={\bf 0}$ of Eq.~(\ref{eq2-1}) now constitutes 
the boundary $r=0$ of Eq.~(\ref{eq2-8})  
(the other one being $r\rightarrow\infty$).
Equation (\ref{eq2-8}) can be rewritten in a form
which is useful for the singularity classification following
Feller \cite{feller55} or Van Kampen \cite{Vankampen} (\ref{appa}),
and to face explicitly the limit case $\xi=2$ (\ref{appb}).
Let us introduce $n(r,t)=\partial_r N(r,t)$ such that $n(r,t)dr$ is the
average number of particles seen by the particle at the origin in the
spherical shell of radius $r$ and infinitesimal thickness $dr$ around it.
Equation (\ref{eq2-8}) in terms of this function becomes
a one-dimensional Fokker-Planck equation
\be 
\partial_t n(r,t)=(a+b)\partial^2_{r}[r^\xi n(r,t)]-(d-1)a
\partial_{r}[r^{\xi-1} n(r,t)]\,,
\label{eq2-8b}
\ee 
which depends explicitly on the dimension $d$ only through
the coefficient $(d-1)$ in the right hand side.

\subsection{The {\em exponent hunter} method for the ``rough'' case $\xi<2$}
We now introduce a new technique for the rough flow case $\xi<2$ through
which we not only recover the above Van Kampen's classification, but
also find the exact small $r$ behavior of all the possible solutions
to Eq.~(\ref{eq2-8}), and therefore to Eq.~(\ref{eq2-1}). 
The smooth case $\xi=2$ is presented in \ref{appb}.

Since $N(r,t)$ is by definition smooth and finite at
small $r$, this method basically amounts to looking for the solution
of Eq.~(\ref{eq2-8}) in the form of the appropriate power expansion
for $N(r,t)$.  Let us assume $\xi<2$ and that $N(r,t)$ at any time can
be expanded at small $r$ in some power series: 
\be N(r,t) = \sum_{l=0}^{\infty}c_l(t)r^{\beta_l}\:\: 
\mbox{with }\beta_i<\beta_j
\mbox{ if }i<j\,,
\label{eq2-9}
\ee
where all $\beta_l$ are independent of $t$. 
Then we construct a recursive solution, by substituting this expression 
into Eq.~(\ref{eq2-8}), 
\bea
\sum_{l=0}^{\infty}\dot c_l(t)r^{\beta_l}&=&
\sum_{l=0}^{\infty}\beta_l\left[(a+b)(\xi+\beta_l-d)\right.\nonumber\\
&+&\left.(d-1)b\right]c_l(t)r^{\xi+\beta_l-2}\,.
\label{eq2-10}
\eea 
The only possibility to have a solution of Eq.~(\ref{eq2-10}),
being $\xi<2$, is that either (i) $\beta_0=0$ or (ii)
$[(a+b)(\xi+\beta_0-d)+(d-1)b]=0$ \footnote{Otherwise the lowest order
power at the right-hand side of the equation cannot be matched by any
term of the left-hand side}.  In order to accept or not these
conditions, we have to take into account the statistical and physical
meaning of $N(r,t)$. This requires $N(r,t)$ has to be finite and
non-negative for all finite $r$ and $t$.  Moreover $\partial_r N(r,t)$
has to be the same non-negative, as the number of particles seen
cannot decrease by increasing the radius of the sphere.  Let us
analyze separately the two cases:

\begin{enumerate}

\item Assume $\beta_0=0$. 
The series~(\ref{eq2-9}) satisfying Eq.~(\ref{eq2-10})
needs to have $\forall n\ge 0$ coefficients such that:
\be
\left\{
\ba{ll}
\beta_n=&n(2-\xi)  \\ \\
\dot c_n(t)=& \gamma_{n}c_{n+1}(t) \\ \\
\gamma_n=&
(n+1)(2-\xi)\{(a+b)[\xi-d+(n+1)(2-\xi)]\\
&+(d-1)b\}
\ea
\right.
\label{eq2-11}
\ee 
This solution corresponds to an expansion into integer powers of
$r^{2-\xi}$.  Moreover, we see immediately that, as $\beta_i<\beta_j$
for $i<j$, $\beta_0=0$ implies $N(r,t)$ to be always finite at finite
$r$ and $t$. Note also that, as $(a+b)>0$ and $\xi<2$, the inequality
$\gamma_i<\gamma_j$ holds for $i<j$ too. Therefore the necessary and
sufficient condition preventing both $N(r,t)$ and $\partial_r N(r,t)$
from becoming negative at sufficiently small $r$ is $\gamma_0\ge 0$,
implying that $\gamma_i\ge 0$ for all integer $i$.  Hence, the
inequality
$$
\gamma_0 \equiv  (2-\xi)[(a+b)(-d+2)+(d-1)b]\ge 0
$$
provides the only condition of acceptability of this solution.  
It can be rewritten in terms of the positive ratio $a/(a+b)$ as 
\be
\frac{a}{a+b}\le {1\over d-1}\,,
\label{eq2-12}
\ee 
i.e., ${\cal P}\ge (d-2+\xi)/(2\xi)$ for the compressibility.
The most important feature of the solution with $\beta_0=0$ is that,
when ${\cal P}>(d-2+\xi)/(2\xi)$ strictly,
the function $N(r,t)$ converges for $r\rightarrow 0$ to a  
positive value $c_0(t)$ growing with time.
This means that particles coalesce more and more in time
forming massive point clusters. Indeed the condition $c_0(t)>0$ means
that the average
conditional particle density $\Gamma(\bm r,t)$ develops a contribution
proportional to the Dirac delta function $\delta(\bm r)$ whose
increasing weight is exactly $c_0(t)$.  From the point of view of
diffusion theory this corresponds either to an {\em adhesive} or 
to a {\em
regular absorbing} boundary at $r=0$ for Eq.~(\ref{eq2-8b}).  The
second important feature is given by the first correction
$c_1(t)r^{\beta_1}$, with $\beta_1=(2-\xi)$, to the leading constant
term of $N(r,t)$. This implies that at sufficiently small $r>0$ the
function\footnote{Or more precisely to its angular average
$(1/\Omega_d)\int d\Omega\,\Gamma(\bm r,t)$.}  $\Gamma(\bm r,t)$ is
proportional to $r^{2-\xi-d}$ with the amplitude proportional to $c_1(t)$.  

Instead for ${\cal P}= (d-2+\xi)/(2\xi)$ we have $\gamma_0=0$
identically.  As a consequence, if $c_0(0)=0$ it will be $c_0(t)=0$
and $N(0,t)=0$ at all $t$. That is $\Gamma(\bm r,t)$ develops no Dirac
delta function contribution and no coalescence occurs. In other words
the probability to find any two particles at vanishing distance is
zero at all time. However, at small $r$ the function $\Gamma(\bm r,t)$
is proportional to $r^{2-\xi-d}$ (i.e. diverging and integrable) which
denotes a simple clustering of the particle distribution.  This
behavior is typical of the singularity $\bm r={\bf 0}$ acting as either
a repulsive entrance boundary, or a reflecting regular boundary.  

\item
Assume now that $(a+b)(\xi+\beta_0-d)+(d-1)b=0$. The 
related solution of Eq.~(\ref{eq2-10}) is given by
the equalities
\be
\left\{
\ba{l}
\beta_0=d-\xi-\frac{(d-1)b}{a+b}=1-\xi+\frac{(d-1)a}{a+b}\\ \\
\beta_n=\beta_0+n(2-\xi)  \\ \\
\dot c_n(t)= \lambda_{n}c_{n+1}(t) \\ \\
\lambda_n=\left[\beta_0+(n+1)(2-\xi)\right]
\left[(n+1)(2-\xi)(a+b)\right]
\ea
\right.
\label{coeff}
\ee 
Again the solution can be found as an expansion in 
integer powers of $r^{2-\xi}$, but there is also a ``singular''
multiplicative contribution $r^{\beta_0}$.  In order to have finite
$N(r,t)$ at finite $r$ the following restriction has to be imposed: 
\[
\beta_0=d-\xi-\frac{(d-1)b}{a+b}\ge 0\,,
\]
which can be recast to
\be
\frac{a}{a+b}\ge \frac{1-\xi}{d-1}\,,
\label{eq2-13}
\ee 
i.e., ${\cal P} \le d/\xi^2$ in terms of compressibility.
This is the only necessary and sufficient condition for the
acceptability of this solution.  In fact, since $\xi<2$, if
$\beta_0\ge 0$ we have from the last of Eq.~(\ref{coeff})
$\lambda_n>0$ for all $n$, and this guarantees both $N(r,t)$ and
$\partial_r N(r,t)$ to be finite and non-negative.  

When $a/(a+b)>(1-\xi)/(d-1)$ (i.e.  ${\cal P} < d/\xi^2$)
strictly, one has $N(0,t)=0$ for all $t$ and no coalescence occurs.
In other words different particles have zero probability at any time
to be found at the same spatial point. In terms of diffusion theory
this is possible only if the point $r=0$ behaves either as a repulsive
entrance boundary or as a regular reflective one. Moreover at
sufficiently small scale $N(r,t)\simeq c_0(t)r^{\beta_0}$ with
$\beta_0=[d-\xi-b(d-1)/(a+b)]>0$. This implies that  $\Gamma(\bm r,t)$ 
is proportional to $c_0(t)r^{\beta_0-d}$. 
Note that the case $a/(a+b)=1/(d-1)$ above is in the present class and in 
fact the exponents coincide for this value of $a/(a+b)$.
The quantity $D=\beta_0$ plays the role of a local (density) fractal
dimension of the particle distribution.

Finally, for $a/(a+b)=(1-\xi)/(d-1)$ (i.e. ${\cal P} =
d/\xi^2$) we have again $\beta_0=0$ and this solution
belongs to the class of solutions with the adhesive or absorbing behavior
at $r=0$ seen above for $a/(a+b)>1/(d-1)$ and developing particle coalescence. 

\end{enumerate}

At this point we can draw the following conclusions for the rough
stochastic flows with $\xi<2$. The physics described by
Eq.~(\ref{eq1}) can be summarized in three different regimes (phases) 
depending on the values of the compressibility parameter
${\cal P}$:\\ 
(i) For ${\cal P}\le (d-2+\xi)/(2\xi)$ (weak compressibility) 
the solution is unique and coincides with the
``repulsive'' solution characterized by
$\beta_0=[d-\xi-b(d-1)/(a+b)]>0$ showing no coalescence but simple
clustering. 
This confirms that for these values of the
compressibility ${\cal P}$, the singularity at $r=0$ acts as a
repulsive entrance boundary: particles never collide.\\ 
(ii) For $(d-2+\xi)/(2\xi)<{\cal P}<d/\xi^2$ (intermediate compressibility)
both attractive and repulsive solutions are possible, and
clearly also all their compositions. This confirms that
the singularity $r=0$ works as a regular boundary
because particles hit one each other in finite time with non-zero
relative velocity.  Accordingly, it is necessary to select an absorbing, 
reflecting or mixed boundary condition to determine a single solution.
In realistic situation, the choiche of appropriate boundary conditions 
is, of course, suggested by physical considerations on the specificity of 
the interactions among pollutant particles in the fluid environment.
Moreover, in real flows, an important role is played by the existence, at small
separation, of two characteristic scales $\ell_{\nu}$, $\ell_{\kappa}$ 
related to purely viscous and diffusive motion respectively. 
Below them, the Kraichnan picture does not hold anymore. 
The attractive solution has to be selected when 
$\{\ell_{\nu},\ell_{D}\}\to 0$ in such a way the Schmidt number 
$\mbox{Sc} =\nu/\kappa$ diverges fast enough \cite{gaw,lejan,vandenE}.\\
(iii) Finally, for ${\cal P}\ge d/\xi^2$ (strong compressibility) the solution
is again unique showing particle coalescence signed by the development
of an increasing delta function at $r=0$ in $\Gamma(\bm r,t)$.
It confirms that for such strong compressibility, $r=0$ works as
an adhesive (or exit) boundary for which particles collide
in finite time but with vanishing relative velocity.\\
It is important to
stress that, due to the structure of the equation system for the
coefficients $c_n(t)$, the only possible stationary solution is given
by $c_0>0$ and $c_n=0$ for $n\ge 1$.  Clearly this is a real
and physically meaningful stationary solution only for the natural
repulsive case at low compressibility ${\cal P}\le (d-2+\xi)/(2\xi)$ or
for the regular boundary case at intermediate compressibility 
$(d-2+\xi)/(2\xi)<{\cal P}<d^2/2\xi$ with reflecting boundary condition.
For the other cases this stationary solution is unphysical amounting to  
placing initially all the system particles at the same spatial point.
It is in fact known from the theory of stochastic processes \cite{Vankampen} 
that when either an adhesive boundary or a regular absorbing boundary 
are present no stationary solution is possible.

\section{Discussion and conclusions}\label{fine}
In this paper we have presented a different approach to analyze and
classify the solutions of the Kraichnan ensemble.  Differently from
the previous methods \cite{verg-gaw,gaw}, our approach focuses on the
equation for the two-point correlation function of the pollutant
density instead of the correlation function of the passive scalar
which is the customary quantity studied in turbulence. The two
equations are known to be equivalent being one the adjoint of the
other.  The reason to suggest this alternative method is
two-fold. First, it represents a more intuitive approach to the
problem based on a natural regularization of the basic diffusion
equation at small scales.
Second, the new method is so general that can be potentially applied to
more complex flows.

The key points of the present approach are: (i) transforming the
fundamental diffusion equation for two-point correlations
$\Gamma(\bm r,t)$ into an equation for the integrated mass of pollutant
$N(r,t)$ surrounding a generic pollutant particle for which the
singularity is smoother; (ii) developing an ``exponent hunter''
technique, consisting in finding the appropriate power series
expansion allowing an exact and quantitative classification of the
particle-particle correlations at finite inter-particle distance.

In this way all the possible behaviors of the small separation
singularity is obtained directly from the explicit solution of the
equation for $N(r,t)$, and the classification becomes straight
forward.

Finally, it is noteworthy to observe that the crossover from intermediate to 
strong compressibility corresponds in multiplicative noise field theories 
to the the non-equilibrium second order transition from an active to an 
absorbing phase.
In particular the adhesive behavior of the singularity for strong 
compressibility Kraichnan models stands for the absorbing phase in field 
theories \cite{multiplicative}.  

\appendix
\section{The Van Kampen's classification}
\label{appa}
In this appendix we apply to Kraichnan ensemble the boundary (or singularity)
classification introduced by Van Kampen in \cite{Vankampen} for
the general one-dimensional Fokker-Planck equation 
\be 
\partial_t
f(r,t)=\frac{1}{2}\partial^2_r[D(r) f(r,t)]- \partial_r[K(r)f(r,t)]\,.
\label{eq3-1}
\ee describing the evolution of the PDF $f(r,t)$ of the position $r$
of a particle at time $t$.  Indeed Eq.~(\ref{eq2-8b}) is exactly of
this type with $D(r)=2(a+b)r^{\xi}$ and $K(r)=(d-1)ar^{\xi-1}$ and
with the singularity to be classified $r=0$.  Van Kampen's
classification for a singularity $r=0$ is based on the analysis of the
behavior for $\eps\rightarrow 0$ of the integrals: 
\be 
\left\{
\begin{array}{l}
L_1=\int_\eps^{r_0}dr\, e^{\phi(r)}\\
L_2=\int_\eps^{r_0}dr\, e^{\phi(r)}\int_{r_0}^r dr' 
\frac{e^{-\phi(r')}}{D(r')}\\
L_3=\int_\eps^{r_0}dr\frac{e^{-\phi(r)}}{D(r)}
\end{array}
\right.
\label{eq3-2}
\ee
where
\[\phi(r)=-2\int_{r_0}^r dr'\frac{K(r')}{D(r')}\]
and with $x,r_0>0$. In general one can show that

\begin{enumerate}

\item If for $\eps\to 0$ we have $L_1\to +\infty$, the singularity $r=0$
behaves as a {\em natural repulsive boundary} as the particle starting
from $r_0>0$ has zero probability to reach $r=0$ \cite{Vankampen}.
This means that starting the motion close to $r=0$, the particle run
away from the singularity never touching it. Consequently,
Eq.~(\ref{eq3-1}) need not an additional boundary condition at
$r=0$ and the solution, once the initial and the other possible
boundary conditions are given, is unique.

In general one
distinguishes two sub-cases depending whether the conditional mean
escape time from the singularity to a finite distance is finite
or infinite. In the first sub-case one can
show that the solution converges in time to a stationary state.  The
conditional mean escape time can be evaluated as follows.  Let us put
at $r=\epsilon>0$ (which is a regular point) a reflective boundary and
start the dynamics from a generic $r_0\in (\epsilon,r_1)$ with
$r_1>\epsilon$. It is possible to show \cite{Vankampen} that the mean
escape time through $r_1$ conditioned to starting the dynamics from
$r_0$ and having a reflecting boundary at $\epsilon$ is
\be
\tau(r_1,\epsilon|r_0)=\frac{\int_{r_0}^{r_1}dr\,e^{\phi(r)}}
{\int_\epsilon^{r}dr'\,e^{\phi(r')}} \,.
\label{eq-tau}
\ee
We have an entrance boundary at $r=0$ if $\tau$ remains finite for
$r\to\epsilon\to 0$. Instead, if $\tau$ diverges, $r=0$ is a proper
natural repulsive boundary. We show below how to distinguish these two
possibilities in our Kraichnan case.

\item If for $\epsilon\to 0$ we have $L_1< +\infty$ and $L_2\to +\infty$, 
the point $r=0$ behaves
as a {\em natural attractive boundary}. The particle starting at
finite $r$ approaches $r=0$ but in an infinite mean time. As above the
solution is unique and no boundary is needed to be fixed.  Differently
from above, no stationary state is reached.

\item If for $\epsilon\to 0$ we have $L_1,L_2< +\infty$ and 
$L_3\to +\infty$, the point $r=0$ behaves as an
{\em adhesive boundary}. The particle starting at
finite $r$ approaches $r=0$ in a finite time but reaches it with
vanishing velocity. Therefore once the particle has reached $r=0$ it
stays there forever. In other words $r=0$ works ``naturally'' as an 
absorbing boundary. Again the solution is unique and no boundary
condition has to be fixed by hand.  Since the probability to find at $t>0$ the
particle at $r=0$ is finite and increases in
time, no stationary state is reached and
$f(r,t)$ develops a Dirac delta function at $r=0$ with a time
increasing coefficient;

\item Finally, if in the same limit $L_1,L_2,L_3< +\infty$, the point
$r=0$ behaves as a {\em regular boundary}. One can now show that the
particle reaches $r=0$ in a finite time but with a non-zero
velocity. Consequently, a solution to the equation is determined once
a boundary condition at $r=0$ is explicitly fixed.  This condition can
be either absorbing, or reflecting, or mixed.  Only if a purely
reflecting condition is fixed the solution runs towards a stationary
state. Otherwise, as in the previous case, a time increasing Dirac
delta contribution appears at $r=0$.

\end{enumerate}

Before applying this classification to the Kraichnan ensemble, it is
useful to remind again that in this case $D(r)=2(a+b)r^{\xi}$,
$K(r)=(d-1)ar^{\xi-1}$ and $r$ is not the position of a single
particle but the relative distance between two system particles, so
that $r=0$ means a collision between these two particles.  The above
classification can be translated as follows:\\ Case 1 is obtained for
$a/(a+b)\ge 1/(d-1)$, i.e., for ${\cal P}\le (d-2+\xi)/(2\xi)$ (weak
compressibility).  Moreover it is simple to show that for $\xi<2$
(rough flow) the conditional mean escape time from $r=0$,
Eq.~(\ref{eq-tau}), for $r_0\to\epsilon\to 0$ is finite and it behaves
as an entrance boundary, while for $\xi=2$ (smooth flow) this time
diverges and $r=0$ behaves as a proper natural repulsive boundary.  In
other words for $\xi<2$ nearby particles almost surely never collide
and get far away from one each other in finite time, while for $\xi=2$
nearby particles again never collide but increase their relative
distance logarithmically.  \\Case 2 is obtained only when $\xi=2$ and
$a/(a+b)< 1/(d-1)$, i.e., ${\cal P}> d/4$ (strong compressibility for
smooth flow).  Now any pair of particles at finite relative distance
almost surely do not collide approaching one each other very slowly.
The difference between the two sub-cases with $\xi=2$ is very
subtle. This is the reason way in the original Feller's classification
they were included in a unique class of ``natural boundaries''. We see
in \ref{appb} how to distinguish these solutions.  \\ Case 3
is obtained when $\xi<2$ and $a/(a+b)\le (\xi-1)/(d-1)$, i.e. ${\cal
P}\ge d/\xi^2$ (strong compressibility).  In this case any two
particles at finite initial relative distance almost surely collide in
a finite time, but with vanishing relative velocity and consequently
coalesce.  \\ Finally case 4 is obtained for $\xi<2$ and
$(\xi-1)/(d-1)<a/(a+b)<1/(d-1)$, i.e., for $(d-2+\xi/(2\xi)<{\cal
P}<d/\xi^2$ (intermediate compressibility).  Now any pairs of particle
almost surely collide in a finite time with non-vanishing relative
velocity. Therefore to fix the solution of the equation one as to
decide if collisions are either completely elastic (reflecting
boundary at $r=0$) or completely inelastic (absorbing boundary at
$r=0$) or intermediate (mixed boundary). This is the only case in
which it is necessary to fix by hand a boundary condition at $r=0$ to
determine the solution of Eq.~(\ref{eq2-1}).

All this qualitative analysis coincides with the one given by \cite{verg-gaw}
through the boundary condition theory of elliptic operators.

\section{The ``smooth'' case $\xi=2$} \label{appb}
We have seen that in Van Kampen's classification  
the solution of the ``smooth'' case $\xi=2$ to our problem is unique 
and corresponds to a singularity at $r=0$ behaving as a natural
(attractive or repulsive) boundary. 
In $d=1$ the problem $\xi=2$ has been solved by 
\cite{Deutsch} which found the solution to Eq.~(\ref{eq1}) for 
$r_0\to 0$. The $d-$dimensional   
case has been extensively studied in \cite{verg2} by analyzing the time 
evolution of the moments of the separation between two system particles. 
Here we give the exact solution for the average conditional 
radial density of particles whose evolution is described by 
Eq.~(\ref{eq2-8b}). This direct solution clarifies the meaning 
of $r=0$ as a natural boundary and the distinction between an 
attractive and a repulsive case.  

First of all we see that any power law function $\eta(t)r^\alpha$ is a
solution of the Eq.~(\ref{eq2-8b}) with $\xi=2$ if $\eta(t)$
satisfies: 
\be 
\eta(t)=\eta(0)e^{\gamma(\alpha) t}
\label{eq-xi2-c}
\ee 
with $\gamma(\alpha)=(\alpha+1)[(a+b)(\alpha+2)-(d-1)a]$.  By the
physical definition of $n(r,t)$ the exponent $\alpha$ needs to be
larger than $-1$ to have always a finite conditional number of
particles $N(r,t)$ in any sphere of finite radius. Therefore
$\gamma(\alpha)>0$ if $a/(a+b)<(\alpha+2)/(d-1)$.  And in particular
for $a/(a+b)<1/(d-1)$ we have $\gamma(\alpha)>0$ for all permitted
$\alpha$. This gives a first insight into the difference between
attractive and repulsive natural boundary at $r=0$. In fact for
$a/(a+b)>1/(d-1)$ (repulsive case) there are initial conditions which
are depleted by the dynamics, while for $a/(a+b)>1/(d-1)$ (attractive
case) all physical initial condition are amplified.  However the fact
that any spatial power law with a coefficient satisfying
Eq.~(\ref{eq-xi2-c}) is always a solution of Eq.~(\ref{eq2-8b}) looks
unphysical because at any time the average particle density in the
infinite volume is not conserved.  This unphysical aspect is due to
the fact that $\hat d(\bm r)$ is divergent on large separations.
Therefore our model has to be interpreted as valid up to an upper cutoff
$L$.

In order to better clarify the nature of the solution of Eq.~(\ref{eq2-8b})
we therefore study the evolution of $n(r,t)$ with initial condition
\be
n(r,0)=\left\{
\begin{array}{ll}
\gamma r^\alpha&\mbox{for }r\le r_c\\ &\\
0&\mbox{for }r> r_c
\end{array} \right.
\label{eq-nr-in}
\ee
where $\alpha>-1$ and $r_c$ is an arbitrary finite scale.

Equation (\ref{eq2-8b}) can be transformed into a more treatable form
performing the change of variable $r=e^u$ and
considering the function $m(u,t)$ defined by the relation
\[m(u,t)=n(r=e^u,t)\frac{dr(u)}{du}=n(r=e^u,t)e^u\,,\]
which conserves the measure.  Directly from Eq.~(\ref{eq2-8b}) it is
simple to show that $m(u,t)$ satisfies the simple Fokker-Planck
equation at constant coefficients typical of ordinary Brownian motion
with diffusion coefficient $2(a+b)$ and constant drift velocity
$-[b+(2-d)a]$ 
\be 
\partial_t m(u,t)=(a+b)\partial^2_u m(u,t)
+[b+(2-d)a] \partial_u m(u,t)\,.
\label{eq-m}
\ee
The initial condition is given by changing variables in Eq.~(\ref{eq-nr-in}):
\be
m(u,0)=\left\{
\begin{array}{ll}
\gamma e^{(\alpha+1) u}&\mbox{for }u\le u_0=\log r_c\\ &\\
0&\mbox{for }u> u_0=\log r_c
\end{array}\right.
\label{eq-nr-in1}
\ee
The solution of Eq.~(\ref{eq-m}) is easily found by considering the  
Fourier transform $\tilde m(q,t)=\int_{-\infty}^{+\infty}du\,m(u,t)e^{-iqu}$
which, substituted in Eq.~(\ref{eq-m}), reads
\be
\tilde m(q,t)=\tilde m(q,0)\exp\{-[(a+b)q^2-i(b+(2-d)a)q]t\}\,.
\label{eq-mq}
\ee
From Eq.~(\ref{eq-nr-in1}), $\tilde m(q,0)$ is given by
\[\tilde m(q,0)=\gamma \frac{e^{(\alpha+1-iq)u_0}}{\alpha+1-iq}\,.\]
Plugging this expression into Eq.~(\ref{eq-mq}) and inverting the Fourier
transform one finds:
\be
m(u,t)=\gamma\exp[(\alpha+1)(u+c_0 t)]
\int_{v_m(u)}^{+\infty}\frac{dv}{\sqrt{2\pi}}e^{-\frac{v^2}{2}}\,
\label{eq-mu}
\ee
where
\bea
&&c_0=b+(2-d)a+(\alpha+1)(a+b)\nonumber \\
&&v_m(u)=\frac{u-u_0+[b+(2-d)a+2(\alpha+1)(a+b)]t}{\sqrt{2(a+b)t}}\nonumber
\eea
We now study the asymptotic of Eq.~(\ref{eq-mu}) by using the following
well known approximations:
\[\int_{v_m}^{+\infty}\frac{dv}{\sqrt{2\pi}}e^{-\frac{v^2}{2}}\simeq\left\{
\begin{array}{ll}
\frac{e^{-\frac{v_m^2}{2}}}{\sqrt{2\pi}v_m}&\mbox{for }v_m\gg 1\\ \\
1&\mbox{for }v_m\ll -1
\end{array}
\right.
\]
Using this and moving back from $m(u,t)$ to $n(r,t)$, we finally find
for $r/r_c\gg \exp[-c_2t+\sqrt{2(a+b)t}]$ 
\be
n(r,t)\simeq\gamma\,r_c^\alpha
\sqrt{\frac{(a+b)t}{\pi}}\left(\frac{r_c}{r}\right)
\frac{\exp\left[-\frac{(\log \frac{r}{r_c}+c_1t)^2}{4(a+b)t}\right]}
{\log\frac{r}{r_c}+c_2t}
\label{eq-nr1}
\ee
with
\[
\left\{
\begin{array}{l}
c_1=b+(2-d)a\\ \\
c_2=b+(2-d)a+2(\alpha+1)(a+b)\,,
\end{array}
\right.
\]
Eq.~(\ref{eq-nr1}) practically says that, in the region of validity of
the approximation, $n(r,t)$ develops a log-normal behavior whose peak
drifts with velocity $-c_1t$.

Instead for $r/r_c
\ll \exp[-c_2t-\sqrt{2(a+b)t}]$ we have simply
\be
n(r,t)\simeq \gamma\, r^{\alpha}\exp\{(\alpha+1)c_0t\} = 
n(r,0) \exp\{(\alpha+1)c_0t\}\,,
\label{eq-nr2}
\ee
i.e., the same amplifying aforementioned behavior for the scale invariant 
initial condition.

By looking at Eq.~(\ref{eq-nr1}) we can appreciate better the meaning 
of ``attractive'' or ``repulsive'' natural boundary behavior at the 
singularity $r=0$.\\
(i) For $c_1>0$, i.e., ${\cal P}>d/4$, the peak of the log-normal  function
shifts to smaller and smaller scales denoting the attractive behavior of the 
point $r=0$;\\
(ii) for $c_1<0$, i.e, ${\cal P}<d/4$, such peaks moves away from the 
singularity signing the repulsive behavior of such singularity.

 

\section*{References}
\begin{thebibliography}{99}
\bibitem{verg-gaw} 
Gawedzki K and Vergassola M 2000 Physica D {\bf 138} 63.

\bibitem{appl1} 
Pruppacher H R and Klett J D 2003,
{\it Microphysics of Clouds and Precipitation} 
(The Netherlands: Kluwer Academic Press)

\bibitem{appl2} 
Okubo A and Levin S A 2001   
{\it Diffusion and Ecological Problems} (New York: Springer New York). 

\bibitem{appl3} 
Dickinson E and Honary F 1986
{\it J. Chem. Soc., Faraday Trans. 2} {\bf 82} 719.

\bibitem{appl4} Sanz-Anchelergues A, Zhabotinsky AM, Epstein IR and 
Mu\~nuzuri A P 2001 
{\it Phys. Rev. E} {\bf 63}, 056124. 

\bibitem{appl5}
Shraiman B I and Siggia E D 2000 {\it Nature} {\bf 405} 639. 

\bibitem{lopez} 
Hern\'andez-García1 E and L\'opez C 2004
{\it Phys. Rev. E} {\bf 70} 016216.

\bibitem{cecco}
Cecconi F, Gonnella G and Saracco G P 2007
{\it  Phys. Rev. E} {\bf 75} 031111.

\bibitem {displa} 
Gabrielli A 2004 {\it Phys. Rev. E} {\bf 70} 066131.

\bibitem{nostro1} 
Gabrielli A and Cecconi F 2007 {\it J. Stat. Mech.} P10007.

\bibitem {Deutsch} 
Deutsch J M 1985 {\it J. Phys. A: Math. Gen.} {\bf 18} 1449.

\bibitem{mehlig} 
Wilkinson M and Mehlig B 2003 {\it  Phys. Rev. E} {\bf 68} 040101.

\bibitem{kraich} 
Kraichnan R H 1968 {\it Phys. of Fluids} {\bf 11} 945.

\bibitem{falkovich-rev} 
Falkovich G, Gawedzki K, Vergassola M 2001 
{\it Rev. Mod. Phys.} {\bf 73} 913.

\bibitem {gaw} 
Gawedzki K and Horvai P 2004 {\it J. Stat. Phys.} {\bf 116} 1247.

\bibitem{reed-simon} 
Reed M and Simon B 1972 {\it Methods of Modern Mathematical Physics} vol~2 
(Academic Press).

\bibitem{Cencio}
Bec J, Cencini M and Hillenbrand R 2007
{\it Physica D} {\bf 226} 11.

\bibitem{Cencio1}
Bec J, Cencini M and Hillenbrand R 2007
{\it Phys. Rev. E} {\bf 75} 025301.

\bibitem{multiplicative}
Hinrichsen H 2000  {\it Adv. in Physics} {\bf 49} 815.

\bibitem{Munoz}
Mu\~noz M A 1998 {\it Phys. Rev. E} {\bf 57} 1377.

\bibitem{Gardiner} 
Gardiner CW 1985 {\it Handbook of Stochastic Methods} (Berlin: Springer).

\bibitem{feller55} 
Feller W 1955 {\it Commun. Pure Appl. Math.} {\bf 8} 203. 

\bibitem{Vankampen} 
Van Kampen N G 1992, {\it Stochastic Processes in Physics 
and Chemistry} (Amsterdam: North-Holland Elsevier).

\bibitem{book} 
Gabrielli A, Sylos Labini F, Joyce M and
Pietronero L 2004 {\it Statistical Physics for Cosmic Structures},
(Berlin: Springer-Verlag).

\bibitem{lejan} 
Le Jan Y, Raimond O 2004 {\it Ann. Probab.} {\bf 32} 1247.

\bibitem{vandenE}
Weinan E and  Vanden-Eijnden E 2000 
{\it Proc. Natl. Acad. Sci. USA} {\bf 97} 8200.

\bibitem{verg2} 
Chertkov M, Kolokolov I and Vergassola M 1998 
{\it Phys. Rev. Lett.} {\bf 80} 512.
 
\end{thebibliography}
\end{document}